\documentclass{czjphys}         
\usepackage{amsfonts,amsmath}
\newcommand{\ii}{\mathrm{i}}

\newcommand{\dd}{\mathrm{d}}
\newcommand{\pd}{\partial}
\newcommand{\hh}{\mathcal{H}}

\newcommand{\e}{\mathrm{e}}
\newcommand{\ket}[1]{\left|#1\right\rangle}
\newcommand{\bra}[1]{\left\langle #1\right|}

\newcommand{\const}{\mathrm{const}}

\newcommand{\tr}{\mathop{\mathrm{tr}}}

\newcommand{\II}{\mathbb{I}}
\begin{document}
\title{Interacting noncommutative solitons (vacua)}  
%
\authori{Corneliu Sochichiu}      \addressi{Bogoliubov Laboratory of Theoretical
Physics\\
Joint Institute for Nuclear Research\\ 141980 Dubna, Moscow Reg.\\
RUSSIA\\Institutul de Fizic\u a Aplicat\u a A\c S,\\ str. Academiei, nr. 5,\\
MD2028 Chi\c sin\u au,\\ MOLDOVA}
\authorii{}     \addressii{}
\authoriii{}    \addressiii{}
\authoriv{}     \addressiv{}
\authorv{}      \addressv{}
\authorvi{}     \addressvi{}
%
\headauthor{C. Sochichiu}            
\headtitle{Interacting NC\ldots}             
\lastevenhead{C. Sochichiu: Interacting NC\ldots} 
\pacs{???}     
\refnum{A}
\daterec{XXX}    
\issuenumber{0}  \year{2001}
\setcounter{page}{1}
\maketitle

\begin{abstract}
We consider the dynamics of two interacting lumps/solitons in a
noncommutative gauge model. We show that equations of motion describing this
dynamics can be reduced to ones of a two-dimensional mechanical system which
is well studied and was shown to exhibit stochastic behaviour.
\end{abstract}

\section{Introduction}
Recent development of string theory has shown that noncommutative (NC) models
can arise in certain limits of String Theory \cite{Seiberg:1999vs}. Moreover
the String Field Theory \cite{Witten:1986cc}, which is the theory of second
quantised string, is based on noncommutative geometry too.

Due to their proximity to String Theory noncommutative theories have many
features in common with it. It is worthwhile to mention the IR/UV mixing
\cite{Minwalla:1999px}, which says that in NC theories there is a
correspondence analogous to string theory between contributions at high
energies with those at low ones. Another common feature is a number of
equivalences/dualities which relate different noncommutative models
\cite{Sochichiu:2000bg,Sochichiu:2000kz}.

This list is continued by the so called noncommutative soliton which was
found in \cite{Gopakumar:2000zd}, in the limit of large noncommutativity
$\theta\to\infty$ and further generalised to finite $\theta$ but nontrivial
gauge field background \cite{Sochichiu:2000rm}. These solitons are believed
to correspond in the string theory language to \emph{branes.}

Noncommutative theories can alternatively be treated as theories of linear
operators defined over a (infinite-dimensional separable) Hilbert space. In
the operator language, noncommutative solitons look like projectors to
finite-dimensional subspaces of the Hilbert space, while the gauge fields
split into parts corresponding respectively to the finite dimensional
subspace and its orthogonal completion \cite{Sochichiu:2000rm}.

In this note we are going to consider gauge field ``solitons'', i.e.
solutions for gauge fields which are projectors to finite-dimensional
subspaces of the Hilbert space rather than ones for the scalar field. For
more details the reader is referred to Ref. \cite{Sochichiu:2001am}.

\section{The Model}
Consider the noncommutative gauge model described by the following action,
\begin{equation}\label{action}
  S=\int\dd t\tr\left(\frac{1}{2}\dot{X}^i\dot{X}^i+\frac{1}{4g^2}
  [X^i,X^j]^2\right).
\end{equation}
Here fields $X^i$, $i=1,\dots,D$ are time dependent Hermitian
operators defined on some separable infinite-dimensional Hilbert
space $\hh$.

Provided the Gauss law constraint $G\equiv[X_i,\dot{X}_i]=0$, the model
(\ref{action}) describes the noncommutative Yang--Mills(--Higgs) model in
the temporal gauge $A_0=0$ in a sense we explain below
\cite{Sochichiu:2000ud,Sochichiu:2000bg,Sochichiu:2000kz}. This appears also
to be the same with the bosonic part of the Hilbert space i.e. $N=\infty$
BFSS M(atrix) model \cite{Banks:1997vh}.

Equations of motion corresponding to this action look as follows,
\begin{equation}\label{EqM}
  \ddot{X}_i+\frac{1}{g^2}[X_i,[X_i,X_j]]=0.
\end{equation}
One may find static classical solution $X_i=p_i$,
\cite{Sochichiu:2000ud,Sochichiu:2000bg,Sochichiu:2000kz}, satisfying,
\begin{equation}\label{ppp}
  [p_i,p_j]=\ii \theta_{ij}^{-1},
\end{equation}
for which we assume irreducibility, i.e. any quantity $F$ commuting with all
$p_i$: $[p_i,F]=0$ have to be proportional to unity operator. In particular,
$\theta^{-1}_{ij}$ should be nondegenerate. This property ensures that any
operator can be formally expressed as an operator function of of the
operators $x^i=\theta^{ij}p_j$, by means of its Weyl symbol. Expanding
fluctuations around this solution, $X_i=p_i + A_i$, and Weyl ordering
operators $A_i$ with respect to $x^i$ one gets precisely the
$(D+1)$-dimensional noncommutative Yang--Mills model for the fields $A_i(x)$.

Weyl symbol $f(x)$ of an operator $f$ can be treated as an
ordinary function subject to the Moyal or star product,
\begin{equation}\label{star}
  f*g(x)=\left.\e^{\ii\theta\epsilon^{ij}\pd_i\pd'_j}
  f(x)g(x')\right|_{x'=x},
\end{equation}
here $f(x)$ and $g(x)$ are Weyl symbols of some operators $f$ and
$g$,  $f*g(x)$ is the Weyl symbol of their product and $\pd_i$,
$\pd'_i$ denote the derivatives with respect to $x^i$ and
${x'}^i$.

Integration of a Weyl symbol corresponds to $2\pi\theta\times$trace of the
respective operator. An important difference of the Weyl--Moyal algebra of
functions defined above from the commutative one (of course, beyond its
noncommutativity) is that the derivative is an internal automorphism of the
algebra. The partial derivative with respect to $x^i$ here corresponds to
the commutator,
\begin{equation}
  \pd_i f(x)=\ii(p_i *f-f*p_i)(x)=[p_i,f](x),
\end{equation}

Getting another solution with a smaller number of independent operators say
$p_\alpha$, $\alpha=1,\dots,p$, expanding and Weyl ordering fluctuations
around it, one gets Yang--Mills--Higgs model in a smaller number of
dimensions equal to  $p$, the number of independent irreducible operators. In
the remaining part of this note we will actually consider the model
corresponding to the expansion around a two-dimensional solution,
\begin{equation}\label{2d}
  [p_1,p_2]=\ii \theta^{-1},
\end{equation}
but one should keep in mind the above equivalence.
\section{Lumps}
In this section we review a completely different class of localised
solutions. In order to do this consider the noncommutative analog of complex
coordinates which is given by oscillator rising/lowering operators $a$ and
$\bar{a}$,
\begin{equation}
  a=\frac{1}{\sqrt{2\theta}}(x^1+\ii x^2),\qquad
  \bar{a}=\frac{1}{\sqrt{2\theta}}(x^1-\ii x^2),\qquad
  [a,\bar{a}]=1,
\end{equation}
and the oscillator basis,
\begin{equation}\label{osc}
  \bar{a}a\ket{n}=n\ket{n},\qquad a\ket{n}=\sqrt{n}\ket{n-1},\qquad \bar{a}\ket{n}=
  \sqrt{n+1}\ket{n+1}.
\end{equation}

There is a solution to \eqref{EqM} different from one discussed in the
previous section which is given by commutative matrices. Due to
commutativity they can be diagonalised by a proper gauge transformation and
represented in the oscillator basis as follows,
\begin{equation}\label{solutionX}
  X^i=\sum_{n=0}^{N}c^i_n\ket{n}\bra{n},
\end{equation}
where $c^i_n$ represents $n$-th eigenvalue of the (finite rank) operator
$X^i$. The rank $N$ of the solution \eqref{solutionX} is called the ``soliton
number''. The simplest one-soliton solutions can be written in the form,
\begin{equation}\label{1solit}
  X_i^{(0)}=c_i\ket{0}\bra{0}.
\end{equation}

\emph{We should warn that the term ``soliton'' applied to the above
configuration is conventional. It is not a soliton in the sense of solutions
to the solitonic equations. Moreover, in this case the solution has zero
energy at rest, however geometrically it is ``noncommutative soliton'' in
the sense of \cite{Gopakumar:2000zd}. In this particular case one also can
call them ``vacuum solitons'' or just ``vacua''.}

In the star-product form operator \eqref{1solit} is represented by
the Weyl symbol
 $$
  X_i(\bar{z},z)=2c_i\e^{-2|z|^2}.
 $$
The soliton shifted along noncommutative plane by a (c-number)
vector $u$ is given by
\begin{equation}\label{1solitu}
  X_i^{(u)}=c_i\e^{-\ii p_\mu u^\mu}\ket{0}\bra{0}
  \e^{\ii p_\mu u^\mu}=c_i\e^{- |u|^2}
  \e^{\bar{a}u}\ket{0}\bra{0}\e^{-a\bar{u}}.
\end{equation}
Its Weyl symbol, correspondingly, is given by
$X_i^{(u)}(z)=2c_i\e^{-|z-u|^2}$. The shifted soliton with
constant $u$ is a solution again. When $u$ becomes time-dependent
one can perform a gauge transformation
\begin{equation}\label{shift}
  X_i\to\e^{\ii p_\mu
  u^\mu(t)}X_i \e^{-\ii p_\mu u^\mu(t)},
\end{equation}
which shifts the soliton back to the centre, but produce a kinetic
term $\dot{u}^2/2$. Thus a single noncommutative soliton moves
like a free \emph{non-relativistic\/} particle. It is also stable
since its energy at rest is zero.

In what follows we are going to analyse the situation when there
is a couple of solitons separated by a distance $u$.

\section{A pair of interacting solitons}

A single noncommutative soliton can be always rotated in the Hilbert space
by a gauge transformation to have the ``polarisation'' $\ket{0}$ and
``orientation'' along $X_1$. When there are two such solitons one can choose
\emph{without loss of generality\/} the ``orientation'' to involve
nontrivially only $X_1$ and $X_2$.

Consider two solitons which are obtained from $c\ket{0}\bra{0}$ by shifts
along the noncommutative plane by respectively $u_{(1)}$ and $u_{(2)}$.
Since the dynamics of the centre is free and can be decoupled by a
time-dependent gauge transformation similar to \eqref{shift}, where
$u=u_{(1)}+u_{(2)}$ is the coordinate of the centre. The coordinates of the
solitons reduced to the centre become $\pm u/2$.

Thus, the configuration we consider looks like,
\begin{subequations}\label{2solit}
\begin{align}\label{2solit1}
  &X_1=cVPV^{-1}\equiv c\ket{-u/2}\bra{-u/2},\\ \label{2solit2}
  &X_2=cV^{-1}PV \equiv c\ket{u/2}\bra{u/2}\\
  &X_i=\const,\qquad i=3,\dots,D,
\end{align}
\end{subequations}
where we introduced the shorthand notations,
\begin{align}\label{V}
  & V=\e^{(\ii/2) p_\mu u^\mu}=
  \e^{\frac{1}{2}(a\bar{u}-\bar{a}u)},\\ \label{P}
  & P=\ket{0}\bra{0}.
\end{align}
The hight $c$ can be absorbed by the rescaling of the coupling and the time,
therefore we put it to $c=1$.

In what follows also $X_i$ $i=1,\dots,D$ will enter trivially in the
equation, so in the remaining part of the paper we simply neglect them and
allow the index $i$ to run in the range $i=1,2$. If we were considering more
than two noncommutative solitons we would have to keep more fields.
\section{Dynamics}

The symmetry of the model allows its further simplification. In
particular, the Gauss law conservation,
\begin{equation}\label{glc}
  \dot{G}=0,
\end{equation}
provided by the unitary gauge invariance, implies that operators $X_i$ are
nontrivial only on a two-dimensional subspace $\hh_{(u)}$ of the Hilbert
space spanned by $\ket{\pm u/2}$. therefore, one can restrict the equations
of motion \eqref{EqM} to $\hh_{(u)}$  assuming thus that $X_i$ are
2$\times$2 matrices acting on it.

One can show \cite{Sochichiu:2001am} that the equations of motion together
with initial conditions are equivalent to the system of two-dimensional
particle moving in the potential $X^2Y^2$,
\begin{subequations}\label{2d_part}
\begin{align}\label{X_sc}
  &\ddot{X}=-\frac{2}{g^2}Y^2X, \\ \label{Y_sc}
  &\ddot{Y}=-\frac{2}{g^2}X^2Y,
\end{align}
which are supplied with the initial conditions,
\begin{equation}\label{ic_XY_sc}
  X|_{t=0}= \e^{-\frac12 |u|^2},\qquad
  Y|_{t=0}=-\sqrt{1-\e^{-|u|^2}}.
\end{equation}
\end{subequations}

This model has a number of interesting features studied in
\cite{Baseian:1979zx,Matinian:1981ji,Medvedev:1985ja%
,Aref'eva:1998mk,Aref'eva:1997es}. The motion resulting from above equations
appears to be \emph{stochastic\/} for all values of $u$ except
$u=\sqrt{\theta}\ln2$. When $u=\sqrt{\theta}\ln2$ the motion is periodic but
unstable, infinitesimal deviation from this position brings it back to
stochasticity.

The solution to initial equations \eqref{EqM} is expressed in
terms of solution $X(t)$ and $Y(t)$ to Eq. \eqref{2d_part} as
follows,
\begin{align}\label{s1}
  & X_1(t,\bar{z},z)=\frac12
  \sigma_0(\bar{z},z)+ X(t)\sigma_3(\bar{z},z)+
  Y(t)\sigma_1(\bar{z},z),\\ \label{s2}
  & X_2(t,\bar{z},z)=\frac12
  \sigma_0(\bar{z},z)+ X(t)\sigma_3(\bar{z},z)-
  Y(t)\sigma_1(\bar{z},z),
\end{align}
where $\sigma_a(\bar{z},z)$ are the Weyl symbols of the
two-dimensional Pauli matrices which form the basis for hermitian
operators acting on $\hh_{(u)}$,
\begin{subequations}\label{sigmas}
\begin{align}\label{sigma1}
  \sigma_1(z,\bar{z})&=\frac{2}{\sqrt{1-\e^{-|u|^2}}}
  \left(\e^{-2|z-\frac{u}{2}|^2}-\e^{-2|z+\frac{u}{2}|^2}\right),\\
\label{sigma2}
  \sigma_2(z,\bar{z})&=\frac{2\ii\e^{-2\bar{z}z}}{\sqrt{1-\e^{-|u|^2}}}
  \left(\e^{\bar{z}u-\bar{u}z}-\e^{-\bar{z}u+\bar{u}z}\right),\\
\label{sigma3}
  \sigma_3(z,\bar{z})&=-\frac{2\e^{-\frac12|u|^2}}{1-\e^{-|u|^2}}
  \left(\e^{-2|z-\frac{u}{2}|^2}+\e^{-2|z+\frac{u}{2}|^2}\right)\\
  \nonumber
  & \qquad +\frac{\e^{-2\bar{z}z}}{1-\e^{-|u|^2}}
  \left(\e^{\bar{z}u-\bar{u}z}+\e^{-\bar{z}u+\bar{u}z}\right),\\
\label{sigm0}
  \II(\bar{z},z)&=\sigma_0(z,\bar{z})=
  \frac{2}{1-\e^{-|u|^2}}
  \left(\e^{-2|z-\frac{u}{2}|^2}+\e^{-2|z+\frac{u}{2}|^2}\right)\\
  \nonumber
  & \qquad -\frac{2\e^{-2|z|^2-\frac12 |u|^2}}{1-\e^{-|u|^2}}
  \left(\e^{\bar{z}u-\bar{u}z}+\e^{-\bar{z}u+\bar{u}z}\right).
\end{align}
\end{subequations}

Qualitatively, the solution \eqref{s1}, \eqref{s2} describes lumps
bouncing (stochastically) around ``points'' $z=0,\pm u/2$ of the
noncommutative plane. In the string theory language the heights
of the lumps can be interpreted as transversal (to the
noncommutative 2-brane) coordinates of two 0-branes.

\bigskip
{\small \textbf{Acknowledgement.} This work was supported by RFBR
grant \#99-01-00190, INTAS grants \#00 262, Scientific School
support grant \# 00-15-96046. My participation to the Conference
was supported by the Votruba--Blokhintsev Czech--Dubna
collaboration grant. (Special thanks also to the Coordinator of
Bogoliubov--Infeld program.)}
\bigskip


\providecommand{\href}[2]{#2}\begingroup\raggedright\endgroup

\end {document}